\documentclass{article}%
\usepackage{amsmath}
\usepackage{graphicx}%
\usepackage{amsfonts}%
\usepackage{amssymb}
\begin{document}

\begin{center}%
\[
\]
{\LARGE Entropy production in a p-n photovoltaic converter}
\end{center}%

\[
\]
$^{(a)}$M. Salis \ , P. C. Ricci and F. Raga

\textit{Dipartimento di Fisica - Universit\`{a} di Cagliari}

\textit{INFM- UdR Cagliari - Cittadella Universitaria , 09042
Monserrato-Cagliari, Italy}%

\[
\]
F. Quarati

\textit{Science Payload and Advanced Concepts Office - ESA/ESTEC,}
\textit{Keplerlaan 1, 2201 AZ, Noordwijk, The \ Netherlands}%
\[
\]

\begin{center}
\textbf{Abstract }
\end{center}

The case of a non-biased (light excited) p-n photo-voltaic (PV) cell working
in a closed circuit with a resistive load is investigated by means of
irreversible thermodynamics in the linear range. To our knowledge, this is the
first time that both electronic processes, such as photo-excitations, carrier
diffusions and recombinations, and electrical processes, such as current
generation and the Joule effect, are considered together in the context of
thermodynamics. An equation for the built-up voltage is obtained where the
recombination resistance of the diode and a collection factor of the
photo-carriers, depending on the thickness of the illuminated cell-side, are
included in a natural way.


\begin{center}%
\[
\]
{\large I. INTRODUCTION}
\end{center}

Photovoltaic (PV) cells are devices capable of photon-to-electrical energy
conversion. \ To date, the most diffuse PV devices are based on (mainly
silicon) p-n junction. The promise of new high efficient devices (dye
sensitized, thermo-photovoltaic, hot carrier, organic) has reinforced the
interest of researchers in this matter$^{1}$. Besides the underlying chemical
and physical processes and the technological features, since the seminal work
by Shockley and Quiesser$^{2}$, the main concern about the utilization of PV
devices is solar energy conversion and the related upper theoretical limit of
conversion efficiency. The efforts made in succeeding in this task brought the
attention of researchers in the thermodynamics of heat exchange between the
black body source (sun) and the heat engine (converter). Most of the works
that addressed this problem were grounded on reversible thermodynamics$^{3}$.
But improvements were obtained by considering the entropy production of PV
conversion$^{4,5}$. Recently, the efficiency problem was addressed by
describing solar energy conversion as an explicitly irreversible process$^{6}%
$. To this end, Onsager's reciprocity theorem was used to investigate
near-equilibrium heat exchange. Despite the aforementioned efforts, literature
lacks a study that explicitly takes into account the thermodynamics of
electronic processes occurring in excited materials (carrier photo-generation,
recombination and diffusion) as well as that of electrical current generation
and the joule effect in a resistive load connecting the two PV-cell sides. For
this reason, it is advisable to dwelling upon this topics since thermodynamics
has an intrinsic generality so that in principle more differently
chemical-physical based PV systems can be covered by the same algorithm
implemented for a specific case. Differently from previously cited works, we
are not interested in the efficiency problem. Thus, the paper is structured as
follows. In Section 2 some essential features of irreversible thermodynamics
are briefly recalled as well as the main results of its application to the p-n
diode. In Section 3, the theory is applied to PV cells and, thus, to PV closed
circuits. The case of the biased PV cell is briefly considered.
\[
\]

\begin{center}
{\large II. ENTROPY PRODUCTION IN A p-n DIODE}

\textbf{A. General}
\end{center}

Before to beginning our investigation of the PV cell, it is best to clarify
some points about irreversible thermodynamics. Let us first consider a
(closed) system which can only exchange energy with its environment. Let this
system also host chemical reactions that convert energies of exciting photons
to heat, the state of reactions being characterized by the actual composition
mixture of photo-sensitive molecules. Any change in the mixture composition
causes a change of an associated entropy $S_{c}$ which will be referred to as
''chemical''. Now, let some exciting photons cause departure from the
equilibrium composition. We suppose that the whole energy of absorbed photons
is used only to change the composition mixture. This process is accompanied by
a change $\delta S_{c}<0$. The excited mixture now returns to the ground state
by reaching its original equilibrium composition after an entropy production
$\delta S_{c}^{\prime}>0$ so that $\delta S_{c}+\delta S_{c}^{\prime}=0$ . But
the (dissipative) reactions bringing the mixture to the equilibrium
composition also cause heat production. If the system were thermally isolated,
the whole conversion process would cause its warming accompanied by an
increase of its entropy. \ If the  system temperature is maintained constant
by heat exchange with a thermal reservoir, its entropy remains unchanged while
increases the entropy of the thermal reservoir. In the stationary
nonequilibrium state the system entropy (when it retains a meaning) is
independent of time. The case of an open system is more complicated. To
explain this point, let us suppose that our system is now capable of releasing
excited molecules to its surroundings and of receiving ground state molecules
in such a way matter conservation is allowed (we are interested in stationary
states). Owing to escaping excited molecules, the negative change of the
chemical entropy is not compensated by the positive one due to the dissipative
reactions within the system. We also have that the heat produced by the
dissipative reactions is released directly to the system's surroundings. Thus,
apparently we have a net negative entropy production in the system.
Nevertheless, the composition mixture is time-independent so that the chemical
entropy does not change. In reality, the chemical entropy does not care about
how the ground state is reached. Actually, positive chemical entropy
production is to be considered due to the flow of matter. In the range of
linear phenomenological laws, the entropy production of chemical processes
$dS_{c}/dt$, has a minimum (compatibly with the system constraints) in the
stationary state ($dS_{c}/dt=0$) according to the minimum rate of entropy
production (MREP) criterion$^{7}$. Thus, if we are interested only in
determining the current flow of the exchanged matter, the stationary condition
could be sufficient. Instead, we could find a valuable tool in variational
calculus if we are interested in characterizing the displacement from
equilibrium of the composition mixture. \ 

To formalize the previous considerations, let us to write the configurational
change of entropy as the internal entropy change \ $d_{i}S$ as in
textbook$^{7}$ %

\begin{equation}
d_{i}S=-\frac{1}{T}\sum_{k=1}^{c}\mu_{k}dn_{k} \label{21form2}%
\end{equation}
where $\mu_{k}$ is the chemical potential of the k-component of a mixture
containing $c$ chemical species and $dn_{k}$ is the corresponding molar
change. \ If at any instant the entropy \ changes as a function of chemical
composition as well as other quantities characterizing the system, it is
possible to write an equation for the rate of entropy production, that is,%

\begin{equation}
\frac{d_{i}S}{dt}=\frac{1}{T}\sum_{l}\Gamma_{l}\upsilon_{l}\;.\label{21form4}%
\end{equation}
In eq. (\ref{21form4}) we take into account that $dn_{k}=\sum_{l}\nu_{kl}%
d\xi_{l}\ ,\ d\xi_{l}\ $stands for the displacement of the l-th reaction,
$\nu_{kl}$\ for the stoichiometric coefficients\ ,\ \ \ $\upsilon_{l}=d\xi
_{l}/dt$, for the flux or velocity of the $l$-reaction and $\Gamma_{l}$ \ for
the affinities, that is${}^{7}$, \ $\Gamma_{l}=-\sum_{k}\nu_{kl}\mu_{k}$. \ It
is assumed \ that near equilibrium the fluxes are linear with respect
\ to\ the affinities, that is, $\upsilon_{l}=\sum_{m}L_{lm}\Gamma_{m}$,\ where
$L_{lm}$ are called the phenomenological coefficients$^{7}$ . As shown by
Onsager$^{8}$, \ based on the time reversal invariance of (microscopic)
mechanical laws, the phenomenological coefficients form a symmetric matrix,
that is $L_{lm}=L_{ml}$. In the case of photo-induced change of the
composition mixture we have%

\begin{equation}
dn_{k}=c_{k}\Phi dt+\sum_{l=1}^{r}\nu_{kl}d\xi_{l}\label{21formYY2}%
\end{equation}
where $\Phi$ stands for the flux of the whole absorbed photons (in suitable
units) and $c_{k}$ for the fraction of photon flux inducing molar change of
the k-component. \ It follows from eqs (\ref{21form2}) that eq. (\ref{21form4}%
) is to be replaced by\ $d_{i}S/dt=-\Phi\Gamma_{\Phi}/T+\sum_{l}\Gamma
_{l}\upsilon_{l}/T$ where $\Gamma_{\Phi}=\sum_{k}c_{k}\mu_{k}$. \ 

Formally, the previous considerations about chemical reactions can be extended
to electronic processes in semiconductors: electron-hole pair
photo-generations play the role of molecule photo-excitations; change of
electron and hole level populations plays the role of composition change of
chemical mixture; electron-hole recombinations play the role of dissipative
reactions. Recombinations can generate photons and heat. We assume that the
semiconductor temperature is independent of time and uniform. Of course, in
the cases of PV cells inserted in closed circuits (open PV systems) part of
the photon energies (that used in photo-excitation processes) can be lost in
the connecting load as Joule heat or as some other electrical works. The flow
of electron in the circuit allows to maintain unchanged the internal entropy
in the stationary state.

\begin{center}
\textbf{B. Entropy production in a p-n diode}
\end{center}

Near equilibrium (but also in most practical cases) the rate equations of
electronic processes can be written in the approximation of non-degenerate
statistics. At the stationary state, densities of conduction band electrons
and valence band holes depend on recombination processes and are independent
of trapping in metastable levels$^{9}$. Recombinations of conduction band
electrons and valence band holes may occur directly, that is, by band-to-band
processes, or indirectly, that is, by trapping in localized recombination
levels. In any case, the departure of free carrier densities from equilibrium
values depends on the affinity$^{10,11}$%

\begin{equation}
\Gamma=\mu_{p}+\mu_{n}=kT\ln\left(  \frac{np}{\overline{n}\overline{p}%
}\right)  \;,\label{21formSK6}%
\end{equation}
$\overline{n}$ and $\overline{p}$ standing for the equilibrium densities. In
the quasi-Fermi level (QFL) formalism, it is$^{11,12,13}$
\begin{equation}
\Gamma=F_{n}-F_{p}\label{formAggX3}%
\end{equation}
where $F_{n}$ and $F_{p}$ stand for the QFL of free electrons and holes
respectively. It can be shown that the rate of all the direct or indirect
recombination processes can be written as $\upsilon=L_{EQ}\Gamma,$ where
$L_{EQ}$\ \ is a phenomenological coefficient accounting for all the processes
involved, and that$^{11}$ $\sum_{l}\Gamma_{l}\upsilon_{l}=\upsilon\Gamma$.
Transport processes driven by electric and diffusion forces contribute to the
entropy production depending on the current of electron ($\overrightarrow
{j}_{n})$ and hole ($\overrightarrow{j}_{p})$ carriers, that is,%

\begin{equation}
\frac{d_{i}S}{dt}=\sigma=-\frac{\overrightarrow{j}_{n}}{T}\cdot grad~\mu
_{n}-\frac{\overrightarrow{j}_{p}}{T}\cdot grad~\mu_{p}%
-\frac{e~\overrightarrow{j}_{n}\cdot\overrightarrow{E}}{T}%
+\frac{e~\overrightarrow{j}_{p}\cdot\overrightarrow{E}}{T}+\frac{\upsilon
\Gamma}{T} \label{form17}%
\end{equation}
where $\mu_{n}$\ and $\mu_{p}$ stand for the chemical potential of free
electrons and holes and \ $\overrightarrow{E}=-grad~V$ for the electric field.
In eq. (\ref{form17}) we suppose that the temperature is uniform in the
semiconductor. The carrier currents are%

\begin{equation}
\overrightarrow{j}_{n}=-D_{n}~grad~n-n~\varkappa_{n}~\overrightarrow{E}~~,
\label{form11}%
\end{equation}%

\begin{equation}
\overrightarrow{j}_{p}=-D_{p}~grad~p+p~\varkappa_{p}~\overrightarrow{E}~~.
\end{equation}
where $\varkappa_{n}$ and $\varkappa_{p}$ \ stand for electron and hole
mobilities respectively. At the stationary state, the continuity equations for
the free carriers are
\begin{equation}
div~\overrightarrow{j}_{n}=-\upsilon~~, \label{form1}%
\end{equation}%

\begin{equation}
div~\overrightarrow{j}_{p}=-\upsilon~~.\label{form2}%
\end{equation}
that is, in the linear range,%

\begin{equation}
D_{n}\nabla^{2}n-\overline{n}\varkappa_{n}\nabla^{2}V-\upsilon=0
\label{form43}%
\end{equation}%

\begin{equation}
D_{p}\nabla^{2}p+\overline{p}\varkappa_{p}\nabla^{2}V-\upsilon=0\label{form44}%
\end{equation}
It is to be pointed out that the continuity equations (\ref{form43} and
\ref{form44}) have been proved to be consistent with the MREP criterion by
considering only band-to-band recombinations$^{10}$. The same calculations can
be repeated by including all the recombination processes whose entropy
production has the simple form $\upsilon\Gamma/T$ as well as by including
photo-excitations 

\begin{center}%
\[
\]
{\large III. ENTROPY PRODUCTION IN A PV-CELL}

\textbf{A. Differential equation for the affinity}
\end{center}

For simplicity's sake, in the following we refer to the diode scheme by
Shockley$^{14}$. Densities and field depend on the x coordinate only, the edge
of the p-side is at $x=0$ and $x_{Tp}$ and $x_{Tn}$ mark the transition region
in the p- and n-side, respectively. By using the Einstein relations, that is,
$eD_{p}/\varkappa_{p}=eD_{n}/\varkappa_{n}=kT$, and by taking into account
that in the approximation dealt with $\partial^{2}\Gamma/\partial x^{2}\approx
kT\left[  \left(  1/\overline{n}\right)  \partial^{2}n/\partial x^{2}+\left(
1/\overline{p}\right)  \partial^{2}p/\partial x^{2}\right]  $, we obtain from
the linear combination of eqs (\ref{form43}) and (\ref{form44})%

\begin{equation}
\frac{\partial^{2}\Gamma}{\partial x^{2}}-kT\frac{L}{D_{p}D_{n}}\left(
\frac{D_{p}}{\overline{n}}+\frac{D_{n}}{\overline{p}}\right)  \Gamma=0~~.
\label{form63}%
\end{equation}
Let us define the free carriers lifetimes as
\begin{equation}
\tau_{p}=\frac{\overline{p}}{kT~L_{EQ}}~~\ \ \ \ \ \ \ \ \ \ \ \tau
_{n}=\frac{\overline{n}}{kT~L_{EQ}}~~, \label{formAggYY63}%
\end{equation}
We introduce the characteristic length $l$ of the affinity function defined by%

\begin{equation}
\frac{1}{l^{2}}=kT\frac{L_{EQ}}{D_{p}D_{n}}\left(  \frac{D_{p}}{\overline{n}%
}+\frac{D_{n}}{\overline{p}}\right)  =\frac{1}{D_{p}\tau_{p}}+\frac{1}%
{D_{n}\tau_{n}} \label{formAggYY66}%
\end{equation}
so that eq. (\ref{form63}) becomes%

\begin{equation}
\frac{\partial^{2}\Gamma}{\partial x^{2}}-\frac{\Gamma}{l^{2}}%
=0~~.\label{form64}%
\end{equation}
As boundary condition for the solution of eq. (\ref{form64}) (in the case of
forward bias) we require only that far from the junction the affinity vanishes
(owing to the equality between electron and hole QFLs$^{14}$). Consequently,
the affinity function is%

\begin{align}
\Gamma\left(  x\right)   &  =\Gamma\left(  x_{Tp}\right)  \exp\left[  \left(
x-x_{Tp}\right)  /l\right]  ~\ \ \ \ \ \ x\leq x_{tp}~~,\label{formZZ64}\\
\Gamma\left(  x\right)   &  =\Gamma\left(  x_{Tn}\right)  \exp\left(
x_{Tn}-x/l\right)  ~\ \ \ \ \ \ x\geq x_{tn}~~.\nonumber
\end{align}
We suppose that the transition region is so small that $x_{Tp}\approx
x_{Tn}\approx s_{p}$ , where $s_{p}$ stands for the thickness of the p-region,
so that $\Gamma\left(  x_{Tp}\right)  =\Gamma\left(  x_{Tn}\right)  =V_{a}$
where $V_{a}$ is the voltage of the biased junction$^{14}$.

\begin{center}
\textbf{B. The case of photo-excitation with carrier diffusion}
\end{center}

In the case of band-to-band semiconductor excitation, any absorbed photons
cause the generation of one electron-hole pair so that $c_{k}=1$
(generalization to the multiple pair generation case is simple) and thus
$\Gamma_{\Phi}=\Gamma$. Equation (\ref{form17}) including photo-excitation becomes%

\begin{equation}
\left(  \frac{d_{i}S}{dt}\right)  _{PV}=\sigma-\frac{\Phi\Gamma}{T}~~.
\label{foto3}%
\end{equation}
In the approximation dealt with the continuity eqs. are%

\begin{equation}
D_{n}\frac{\partial^{2}}{\partial x^{2}}n-\overline{n}\varkappa_{n}%
\frac{\partial^{2}}{\partial x^{2}}V-\left(  \upsilon-\Phi\right)  =0~~,
\label{fotoAgg1}%
\end{equation}%

\begin{equation}
D_{p}\frac{\partial^{2}}{\partial x^{2}}p+\overline{p}\varkappa_{p}%
\frac{\partial^{2}}{\partial x^{2}}V-\left(  \upsilon-\Phi\right)  =0~~,
\label{fotoAgg2}%
\end{equation}
Equations (\ref{fotoAgg1}) and (\ref{fotoAgg2}) can be combined to give%

\begin{equation}
\frac{\partial^{2}\Gamma}{\partial x^{2}}-\frac{1}{l^{2}}\left(
\Gamma-\frac{\Phi}{L_{EQ}}\right)  =0~~.\label{foto8}%
\end{equation}
Note that in the linear approximation, that is, by putting $u=kT~\left(
p-\overline{p}\right)  /\overline{p}\approx\Gamma$, eq. (\ref{foto8}) becomes
similar to eq. (10) of ref. [$15$] (in which a general relationship between
dark carrier distribution and photo-carrier collection in solar cell is
presented) provided $\overline{p}$\ is independent of spatial coordinates.
Now, we suppose that monochromatic radiation is used to excite the PV-cell at
the edge of the p-side ($x=0$). Thus, eq. (\ref{foto8}) can be solved by
assuming that the $\Phi\left(  x\right)  $ is ruled by the Lambert-Beer law,
that is,
\begin{equation}
\Phi\left(  x\right)  =\Phi_{0}\exp\left(  -x/d\right)  ~~,\label{foto9}%
\end{equation}
\ $d$ standing for the characteristic penetration length of radiations.
Actually, $d$ can be of the same order of $l$ so that we should account for of
the different properties of the solution of eq. (\ref{foto8}) in the cases of
$d$  smaller, larger or equal to $l$ (this is important if one is interested
in investigating the wavelenght dependence of PV efficiency). To simplify
calculations, let us assume that $d<<l$. With this approximation the solution
of eqs. (\ref{foto8}) can be written as%

\begin{equation}
\Gamma\left(  x\right)  =\left(  \Gamma_{0}+\frac{d^{2}}{l^{2}}\frac{\Phi_{0}%
}{L_{EQ}}\right)  \exp\left(  -x/l\right)  -\frac{d^{2}}{l^{2}}\frac{\Phi_{0}%
}{L_{EQ}}\exp\left(  -x/d\right)  +V_{a}\exp\left(  -\left|  x-s_{p}\right|
/l\right)  \label{foto21}%
\end{equation}
where, in general, $\Gamma_{0}$ and $V_{a}$ are not independent of each other
(now $V_{a}$ cannot be considered the junction voltage). For $x=0$ we have
$\Gamma\left(  0\right)  =\Gamma_{0}+V_{a}\exp\left(  -s_{p}/l\right)  $\ . Of
course, if \ $s_{p}>>l$ only the interaction with photons allow for
$\Gamma\left(  0\right)  \neq0$. We point out that the value of $\Gamma_{0}$
is fixed by the actual net rate of carrier generation in the PV-cell, that
is,
\[
G=\int\left(  \Phi-\upsilon\right)  d^{3}x=AL_{EQ}\int\left(  \frac{\Phi
}{L_{EQ}}-\Gamma\right)  dx
\]
$A$ being the cross section area of the diode. In the case of the open
PV-circuit, at the steady state $G=0$ so that $\left(  d_{i}S/dt\right)
_{PV}=0$ holds. To better illustrate this point, we assume that the parameters
of the affinity function are independent of x (in reality they can change when
passing from the p- to n-region as well as near semiconductor surface).
Moreover, we assume that the thickness of the diode (but not of the p-region)
is much larger than $l$ . Thus, by putting $\Theta_{l}=Al$ and $\Theta_{d}%
=Ad$\ \ we obtain (terms of the order $\left(  d/l\right)  ^{2}$ are disregarded)%

\begin{equation}
G=\left(  \Theta_{d}\Phi_{0}-\Theta_{l}L_{EQ}\Gamma_{0}\right)  ~~.
\label{foto22}%
\end{equation}
Now, it is easy to prove that at the open circuit condition $j_{n}=j_{p}$ so
that, by taking into account that $\left(  \mu_{n}+\mu_{p}\right)  =kT~\ln
np/\overline{n}\overline{p}=\Gamma$,%

\begin{equation}
\left(  \frac{d_{i}S}{dt}\right)  _{PV}=-\frac{j_{p}}{T}\frac{\partial
}{\partial x}\Gamma+\frac{\upsilon\Gamma}{T}-\frac{\Phi\Gamma}{T}~~,
\label{fotoAgg21}%
\end{equation}
whose integral over the whole PV cell volume is (see Appendix )%

\begin{equation}
\Xi_{PV}=\int_{cell}\left(  dS_{i}/dt\right)  _{PV}d^{3}x=\frac{1}{T}%
\Gamma_{0}G=0~~. \label{AggRev1}%
\end{equation}

\begin{center}
\textbf{C. Entropy production in the case of closed photo-voltaic circuit }
\end{center}

Let an electric load be matched to the two diode sides, and the PV-cell
operate in photo-voltaic mode (no applied bias). In this case, a fraction of
photo-generated holes sustains the electric current $I_{ext}$ so that the rate
of recombinations in the semiconductor turns out to be less than that of
photo-generations (thus we are dealing with an open system). In this
condition, the carriers diffusing towards the junction contain an excess of
electrons with respect to the holes, such that it allows the generation of
$I_{ext}$. On this ground, the external current is%

\begin{equation}
I_{ext}=G=e\left(  \Theta_{d}\Phi_{0}-\Theta_{l}L_{EQ}\Gamma_{0}\right)
\label{AggRev2}%
\end{equation}
The condition $I_{ext}\neq0$ implies that negative and positive terms in eq.
(\ref{foto3}) are now imbalanced. To make this clear it is convenient to write
the equation of entropy production in such a way that the external current
will appear explicitly. Thus, by taking into account that
\begin{equation}
j_{n}-j_{p}=\frac{I_{ext}}{eA}~~, \label{foto23}%
\end{equation}
it follows from eq. (\ref{foto3})%

\begin{equation}
\left(  \frac{d_{i}S}{dt}\right)  _{PV}=-\frac{1}{T}~j_{p}\frac{\partial
\Gamma}{\partial x}+\frac{\upsilon\Gamma}{T}-\frac{\Phi\Gamma}{T}-\frac{1}%
{T}\frac{I_{ext}}{eA}\frac{\partial\mu_{n}}{\partial x}~~, \label{foto24}%
\end{equation}
Form the integration of eq. (\ref{foto24}) we obtain (see Appendix )%

\begin{equation}
\Xi_{PV}=\int_{cell}\left(  dS_{i}/dt\right)  _{PV}d^{3}x+\Xi_{En}%
=-\frac{1}{T}\frac{I_{ext}}{e}\Gamma\left(  x_{Tn}\right)  +\Xi_{C}%
\label{foto25}%
\end{equation}
where we also considered the contribution  $\Xi_{C}$ due to the carrier flow
by the external circuit. Another way to obtain eq. (\ref{foto25}) is to
consider the entropy change due to current $I_{ext}$ crossing the junction
against the built up PV-cell voltage (charge separation in the photovoltaic
process) $V=\Gamma\left(  x_{Tn}\right)  =F_{n}\left(  x_{Tn}\right)
-F_{p}\left(  x_{Tn}\right)  ^{14}$. The field at the junction is
$\overrightarrow{E}=gradV$ so that (see eq. \ref{foto3}), $-\left(
\overrightarrow{j}_{n}-\overrightarrow{j}_{p}\right)  \cdot\overrightarrow
{E}/T=\left(  I_{ext}/Ae\right)  \partial V/\partial x/T$ \ which after
integration gives $-\left(  I_{ext}/Te\right)  V$. Now, since $\Gamma\left(
x_{Tn}\right)  \approx\Gamma\left(  s_{p}\right)  \approx\Gamma_{0}\exp\left(
-s_{p}/l\right)  $ it follows from eq. (\ref{foto25})%

\begin{equation}
\Xi_{PV}=-\frac{I_{ext}\Gamma_{0}\exp\left(  -s_{p}/l\right)  }{eT}+\Xi
_{C}~~.\label{fotoXX25}%
\end{equation}
As for $\Xi_{C}$ we may calculate it by taking into account that electrons
leave the n-side at the chemical potential $\left(  \overline{\mu}_{n}\right)
_{n}$\ and holes leave the p-side (electrons from the conductor) at the
chemical potential $\left(  \overline{\mu}_{p}\right)  _{p}$ so that,
analogously to recombination processes, the \ associated affinity can be
written as%

\begin{equation}
\Lambda=\left(  \overline{\mu}_{n}\right)  _{n}+\left(  \overline{\mu}%
_{p}\right)  _{p}\label{AggRev3}%
\end{equation}
which is nothing but the difference in QFLs of the two cell sides. Thus%

\begin{equation}
\Xi_{C}=\frac{\Lambda}{T}\frac{I_{ext}}{e}\label{AggRev4}%
\end{equation}
The affinity $\Lambda$ is actually determined by the constraints due to the
connected external circuit. In the case of a ourely resistive circuit
$\Lambda=e\left(  R+R_{s}\right)  I_{ext}$ where $R$ stands for the resistance
of the connected load and $R_{s}$ for a resistance accounting for interface
effects at the load contacts.
\begin{equation}
\Xi_{PV}=-\frac{\Gamma_{0}\exp\left(  -s_{p}/l\right)  }{eT}I_{ext}%
+\frac{\Lambda}{T}\frac{I_{ext}}{e}~~.\label{formXX23}%
\end{equation}
Note that $RI_{ext}^{2}/e$ also corresponds to the energy lost  in the load
(Joule heat) by the PV system . Now, by taking into account of eq.
(\ref{AggRev2}) we obtain at the stationary state%

\begin{equation}
I_{ext}=\frac{e\Theta_{d}\Phi_{0}}{R+R_{s}+\exp\left(  -s_{p}/l\right)
/e^{2}\Theta_{l}L_{EQ}}\frac{\exp\left(  -s_{p}/l\right)  }{e^{2}\Theta
_{l}L_{EQ}} \label{form65}%
\end{equation}
Equation (\ref{form65}) can be written in a simplest form by putting%

\begin{equation}
I_{g}=e\Theta_{d}\Phi_{0}~~, \label{form66}%
\end{equation}%

\begin{equation}
R_{0}=\frac{\exp\left(  -s_{p}/l\right)  }{e^{2}\Theta_{l}L_{EQ}}=\exp\left(
-s_{p}/l\right)  \frac{\rho_{EQ}}{e^{2}\Theta_{l}}=\exp\left(  -s_{p}%
/l\right)  R_{EQ}~~, \label{form67}%
\end{equation}
Thus, we have%
\begin{equation}
I_{ext}=\frac{R_{0}}{R+R_{s}+R_{0}}I_{g} \label{from70}%
\end{equation}
As for the meaning of $R_{0}$ it is convenient to rewrite it as%

\[
R_{0}=\exp\left(  -s_{p}/l\right)  \frac{\rho_{EQ}}{e^{2}\Theta_{l}}%
=\exp\left(  -s_{p}/l\right)  R_{EQ}%
\]
$\rho_{EQ}=1/L_{EQ}$ being \ the recombination resistance (note that here the
symbol is different with respect to the ones used for electrical
resistances)$^{11,12}$. $R_{EQ}$ can be defined as the electrical
recombination resistance normalized with respect to the volume having a
thickness equal to the diffusion length. Now, the open circuit voltage turns
out to be ($R=\infty$)%

\begin{equation}
V_{\infty}=R_{0}I_{g}=\exp\left(  -s_{p}/l\right)  R_{EQ}I_{g}
\label{formAgg67}%
\end{equation}
Actually, eq. (\ref{form67}) shows that if the thickness $s_{p}$ is very large
with respect to the diffusion length, resistance $R_{0}$ vanishes. This is a
consequence of the increase of the time available for recombinations of
photo-generated carriers before they reach the junction.%

\[
\]

\begin{center}
\textbf{D. The case of a biased PV cell}
\end{center}

The case of biased junction is more concealed since $\Gamma_{0}$ and $V_{a}$
are not independent. Indeed, by using eq. (\ref{foto21}), eq. (\ref{AggRev2})
is to be replaced by%

\begin{equation}
I_{ext}=e\Theta_{d}\Phi_{0}-e\Theta_{l}L_{EQ}\Gamma_{0}-2\left[  1-\exp\left(
-s_{p}/l\right)  \right]  e^{2}\Theta_{l}L_{EQ}V_{a}-e^{2}\Theta_{l}%
L_{EQ}V_{a}\exp\left(  -s_{p}/l\right)  \label{AggRev5}%
\end{equation}
which only fixes a relation between the unknown variables $I_{ext}$ ,
$\Gamma_{0}$ and $V_{a}$.\ The stationary condition can only fix an additional
relation thus leaving under-determined the problem. About the eq.
(\ref{AggRev5}), note that as the thickness of the p-region becomes much
larger than the diffusion length, all the photo-generated carriers recombine
before the junction is reached. Thus, in this case, the electrical current is
due only to the applied bias (diode mode), that is (note that the sign
convention used for the current in the PV cells is the opposite with respect
to the one used for diodes), $I_{ext}=-2e^{2}\Theta_{l}L_{EQ}V_{a}%
=-2V_{a}/R_{EQ}=-I_{s}\left(  eV_{a}/kT\right)  $ which in the linear range is
the current of the biased diode where $I_{s}=2kT/eR_{EQ}$ is the saturation
current$^{11}$. We conjecture that, in the linear range, $V_{a}=0$ when
$\Gamma_{0}$ can match the external constraints, that is, when $\Gamma_{0}%
\exp\left(  -s_{p}/l\right)  =V_{b}+\Lambda$ where $V_{b}$ stands for the
voltage of the biasing battery. Thus, at the condition $I_{ext}=0$ we have
$V_{b}=R_{0}I_{g}=V_{\infty}$.

\begin{center}%
\[
\]
\textbf{Conclusions}
\end{center}

We investigated the PV conversion by means of irreversible thermodynamics.
This has been made by considering explicitly photo-excitation, recombination
and diffusion of carriers in the non-biased PV-cell as well as electric
current and the Joule effect in a resistive load closing the PV-circuit. An
equation for built up voltage and current in a closed PV-circuit is obtained
where a collection factor and a recombination resistance are included.

\begin{center}%
\[
\]
{\large APPENDIX }
\end{center}

Equation (\ref{AggRev1}) can be proved \ after direct and tedious calculations
by taking into account that, in the case of non-biased PV cell ($V_{a}=0$),
eq. (\ref{fotoAgg1}) can be simplified as%

\begin{equation}
D_{p}\frac{\partial^{2}}{\partial x^{2}}\delta p=\left(  \upsilon-\Phi\right)
\label{eqApp0}%
\end{equation}
where $\delta p=p-\overline{p}$ (note that, at the equilibrium, the change of
carrier densities at the junction are connected to the built up field).
Eq(\ref{eqApp0})\ which (see eq. \ref{foto21}) has the solution%

\begin{equation}
p=\overline{p}+\frac{L_{EQ}}{D_{p}}\left\{  \left(  \Gamma_{0}+\frac{d^{2}%
}{l^{2}}\frac{\Phi_{0}}{L_{EQ}}\right)  l^{2}\exp\left(  -x/l\right)
-\frac{\Phi_{0}}{L_{EQ}}\left(  1+\frac{d^{2}}{l^{2}}\right)  d^{2}\exp\left(
-x/d\right)  \right\}  ~~. \label{eqApp1}%
\end{equation}
In the case of open circuit, all photo-generated carriers recombine in the
semiconductor ($G=0$) so that \ $\int_{cell}\left(  dS_{i}/dt\right)
_{PV}d^{3}x=0$. In the case of a closed circuit (I$_{ext}\neq0$) we obtain
from integration of eq. (\ref{foto24})
\begin{align*}
\int_{cell}\left(  dS_{i}/dt\right)  _{PV}d^{3}x  &  =\frac{1}{T}\Gamma
_{0}\left(  \Theta_{l}L_{EQ}\Gamma_{0}-\Theta_{d}\Phi_{0}\right)  -\frac{1}%
{T}\frac{I_{ext}}{eA}\int\frac{\partial\mu_{n}}{\partial x}d^{3}x=\\
&  =-\frac{1}{T}\frac{I_{ext}}{eA}\Gamma_{0}-\frac{1}{T}\frac{I_{ext}}%
{e}\left[  \mu_{n}\left(  \infty\right)  -\mu_{n}\left(  0\right)  \right]
\end{align*}
where $\mu_{n}\left(  \infty\right)  $ is the chemical potential of electrons
at the end-side of the n-region. By taking into account that in the p-region
$\mu_{p}\approx\overline{\mu}_{p}\approx\mu_{p}\left(  s_{p}\right)  $ (holes
are majority carriers), $\mu_{n}\left(  s_{p}\right)  \approx\mu_{n}\left(
\infty\right)  =-\mu_{p}\left(  \infty\right)  $ (equality of QFLs at the edge
of n-side$^{15}$) and that $\mu_{p}\left(  s_{p}\right)  -\mu_{p}\left(
\infty\right)  \approx\Gamma\left(  s_{p}\right)  $ (in the n-region affinity
is determined by the diffusion process of holes through the junction), we have
$\mu_{n}\left(  \infty\right)  -\mu_{n}\left(  0\right)  =\left\{  \left[
\mu_{n}\left(  \infty\right)  +\mu_{p}\left(  s_{p}\right)  \right]  -\left[
\mu_{n}\left(  0\right)  +\mu_{p}\left(  s_{p}\right)  \right]  \right\}
=\Gamma\left(  s_{p}\right)  -\Gamma_{0}$. Therefore%

\[
\int_{cell}\left(  dS_{i}/dt\right)  _{PV}d^{3}x=-\frac{1}{T}\frac{I_{ext}%
}{eA}\Gamma\left(  s_{p}\right)  ~~.
\]
which represents only a part of the entropy production in the PV-cell since it
lacks the contribution due to the carrier flow in the circuit.
\[
\]
{\large REFERENCES}

$^{(a)}$masalis@vaxca1.unica.it

$^{1}$M. A. Green: Physica E \textbf{14}, 11 (2002)

$^{2}$W. Shockley and H. J. Queisser: J. Appl. Phys. \textbf{32}, 510 (1961).

$^{3}$P.T. Landsberg, V. Badescu, in: S. Sieniutycz, A. de Vos (Eds.),
Thermodynamics of Energy conversion and Transport, Springer, New York, 2000,
p. 72

$^{4}$A. Loque and A. Mart\`{\i}, Phys. Rev. B \textbf{55},6994 (1997).

$^{5}$P. Wurfel, Physica E \textbf{14}, 18 (2002)

$^{6}$T. Markvart, P. T. Landsberg, Physica E \textbf{14}, 71 (2002).

$^{7.}$I$.$Prigogine, Introduction to Thermodynamics of Irreversible Processes
, (John Wiley\&Sons, Interscience, N. Y., 1971).

$^{8}$L. Onsager, Phys. Rev., \textbf{37},405 (1931); \textbf{38},2265 (1931).

$^{9}$A. Delunas, V. Maxia and S. Serci: Lettere al Nuovo Cimento \textbf{34},
559 (1982)

$^{10}$V. Maxia, Phys. Rev. B \textbf{25}, 4196 (1982); Phys. Rev. B,
\textbf{21} 749 (1980); Phys. Rev. B \textbf{17}, 3262 (1978).

$^{11}$M. Salis, P. C. Ricci and F. Raga, \textit{Thermodynamic basis of the
concept of recombination resistance}, submitted to Phys. Rev. B.; ArXiv:cond-mat/0503021

$^{12}$W. Shockley and \ W. T. Read, Jr. , Phys. Rev. \textbf{87} 835 (1952).

$^{13}$ W. G. Pfann and W. van Roosbroek, J. Appl. Phys. \textbf{25}, 1422 (1954)

$^{14}$W. Shockley: Electrons and Holes in Semiconductors (Van Nostrand,
Princeton, 1950) Ch. 12.

$^{15}$M. A. Green, J. Appl. Phys. \textbf{81}, 1 (1997)
\end{document}